%% file: main.tex
\documentclass[CTAC,short]{cta-v2mod}

\usepackage{nicefrac}					
\usepackage{amssymb}					
\usepackage{float}						
\usepackage{lineno}						
\usepackage{multirow}					
\usepackage{subcaption}					
\usepackage{textcomp}					
\usepackage{array}						
\usepackage{url}
\usepackage{pdflscape}						
\usepackage[backgroundcolor = red]{todonotes}	
\usepackage{textgreek}					
\usepackage{color, soul, xcolor}				
\usepackage{listings}	
\usepackage{pdfpages}

\usepackage[utf8]{inputenc}

\usepackage{graphicx}
\usepackage{amsmath,amssymb}
\usepackage{aas_macros}

\definecolor{officegreen}{rgb}{0.0, 0.5, 0.0}

\title{Probing Dark Matter and Fundamental Physics with the Cherenkov Telescope Array\\
}

\graphicspath{{./},{./figures}}

\author{The CTA Dark Matter and New Physics working group for the CTA consortium}
\date{April 2021}

\begin{document}

\includepdf[pages=-]{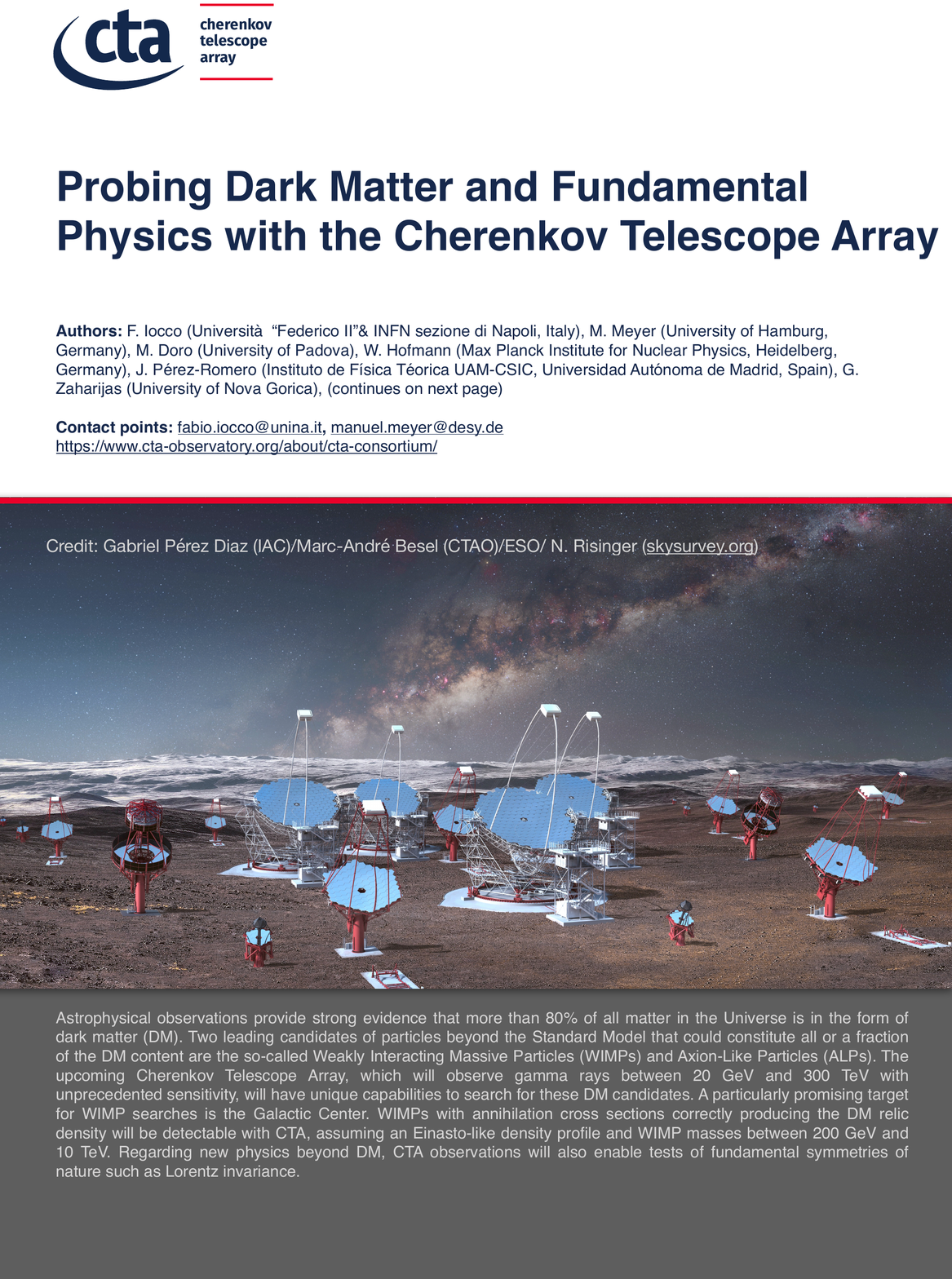}


\input{author_list}


\chapter{Introduction}
Astrophysical observations of different objects such as spiral galaxies, galaxy clusters, elliptical galaxies, galaxies with low surface brightness and dwarf spheroidal galaxies 
all indicate the existence of a non-baryonic dark component of matter~\cite{Bertone_2018}.
Together with independent observations of the Early Universe 
this is interpreted in the self-consistent paradigm 
of a $\Lambda$ cold dark matter Universe~\cite{Planck2020}.
No mundane astrophysical population, and no particle predicted by the Standard Model, can provide a ``standard physics''  candidate for the dark matter (DM), since they do not fulfill the requirements of the above-mentioned observations~\cite{Bertone_2005}.
Thus, evidence for physics beyond the Standard Model of particles intriguingly arises from astrophysical observations.

Many well-motivated extensions of the Standard Model have been devised, which provide new particle candidates that, on the one hand, comply with the observational requirements for DM, 
and, on the other hand, provide a minimal coupling to the visible sector, offering a window for potential observations.
Interestingly, for many of the potential candidates the coupling with Standard Model particles would produce radiation at energies above the GeV scale, thus making gamma-ray observatories ideal tools in the search for the dark matter. Such radiation would be visible ``in excess'' on the top of the standard astrophysical sources, thus requiring a rigorous characterization of the fore and background signals, and a strategic choice of target objects.
Weakly interacting Massive Particles (WIMPs) and Axion Like Particles (ALPs) are recognized to be particularly promising class of candidates \cite{Bertone_2005, Irastorza_2018}. 

With a factor 5-20 higher sensitivity, better spectral and angular resolution than current Imaging Atmospheric Cherenkov Telescopes \cite{2019APh...111...35A}, the Cherenkov Telescope Array (CTA) will be the prime instrument for gamma-ray astronomy and DM searches over the coming decades.
CTA will be a unique tool for the astrophysics community to delve into the exploration of physics beyond the Standard Model.

\chapter{WIMP searches}
\section{Rationale of WIMP annihilation / mechanism}
Weakly Interacting Massive Particles are promising particle candidates for the DM: the same process of self-annihilation into Standard Model particles, which dictates their relic abundance in the Early Universe would allow them today to produce prompt or secondary gamma-ray emission in environments rich in DM.
Such particles with masses and couplings at the electroweak scale would be a compelling solution to the DM puzzle because their existence would explain the presently measured DM abundance as a result of the thermal history of the Universe
and, at the same time, it might address the naturalness problems in the Standard Model of particle physics.

CTA is expected to reach the benchmark value of the (velocity weighted) annihilation cross-section of roughly ($3\times 10^{-26}$ cm$^3$ s$^{-1}$) for WIMPs produced thermally in the early
Universe, thus being the only gamma-ray instrument 
--planned or existing-- capable of testing physics beyond the Standard Model at the TeV scale in astrophysical environments. 
This complements direct DM searches in underground laboratories or WIMP searches at colliders (see, e.g., Ref. \cite{Balazs:2017hxh}).

\section{Targets}
CTA will observe different targets in the search for DM, which are selected on the rationale that the signal produced by DM annihilation is larger in close-by regions with high DM density.

\begin{figure}[t]
\centering\includegraphics[width=0.8\linewidth]{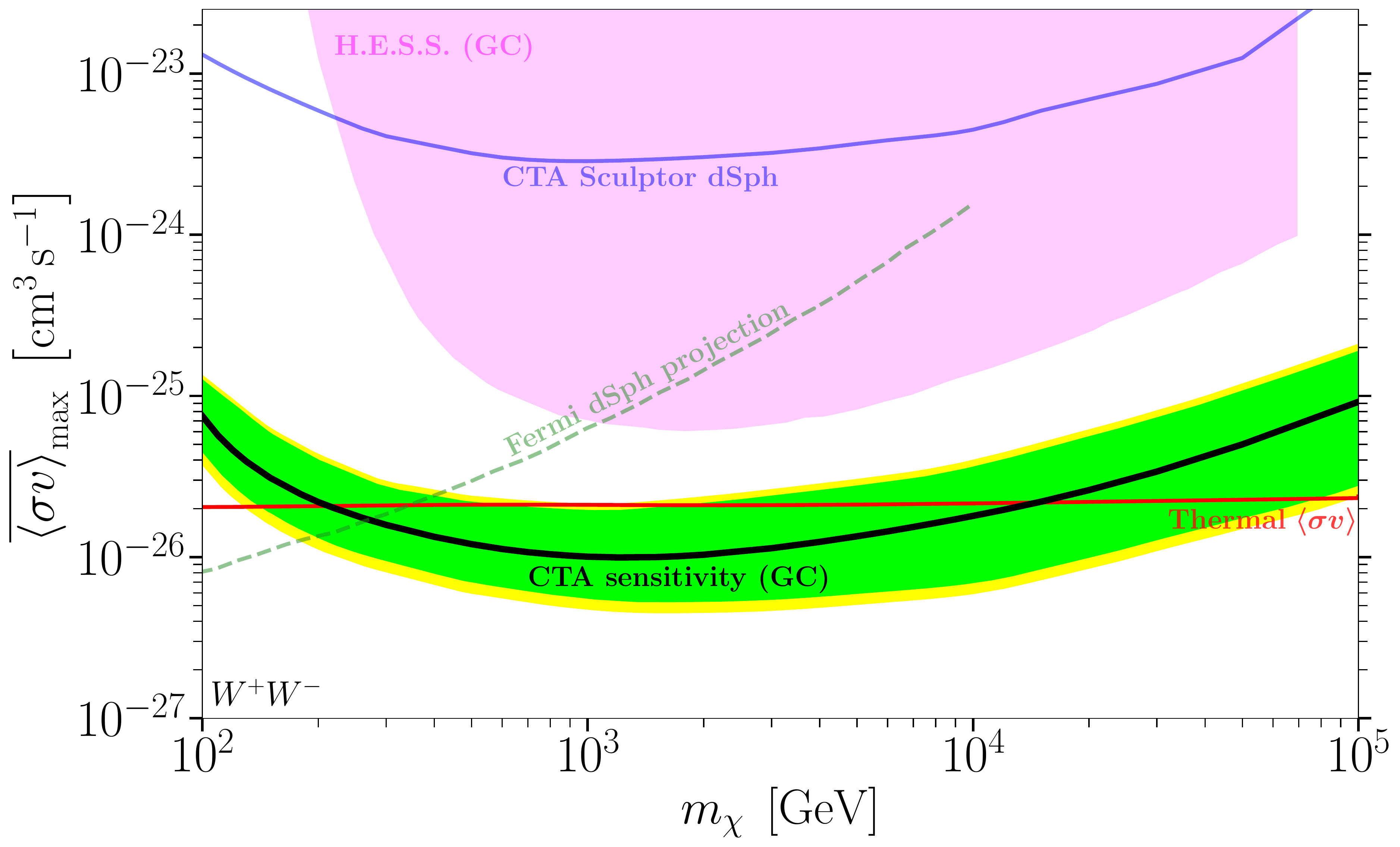}
\caption{The CTA sensitivity curves derived in Ref.~\cite{Acharyya:2020sbj} for annihilation into $W^+W^-$.
Between 200\,GeV and 15\,TeV, CTA observations of the GC will probe the DM annihilation cross section which results in the correct relic abundance (black line and green and yellow shaded regions for median, 68\,\%, and 95\,\% sensitivity estimates from Monte Carlo simulations, respectively), shown by the red line  \cite{Bringmann_2018}. 
Note that the projected sensitivity of CTA shown here~\cite{Acharyya:2020sbj} includes estimates of the expected systematic uncertainties.
Also shown are projections for 
Fermi-LAT observation of dSph  galaxies where dSphs discoveries with the Rubin Observatory are taken into account (dashed green line) and H.E.S.S. observations of 
the GC (purple region) \cite{Abdallah:2016ygi}.
Projections for CTA from the observation of one dSph are shown in blue ~\cite{CTAConsortium:2018tzg}.
\label{fig:summary-sensitivity}
}
\end{figure}

\paragraph{Galactic center (GC).}
The GC region has the largest DM signal (albeit large uncertainties), of all known targets and will be an essential DM target for CTA. However, the GC hosts a rich environment of astrophysical gamma-ray emitters, resulting in complex backgrounds for DM searches. In the recent  CTA Consortium publication \cite{Acharyya:2020sbj} the collaboration presented the most detailed assessment of the CTA sensitivity to DM signals at the GC thus far.
From the perspective of fundamental physics, the most important finding of this work is the confirmation that CTA will be able to reach the thermal cross-section for TeV-scale DM--a milestone for probing the WIMP paradigm at these masses--for a large range of well-motivated assumptions about the instrument’s performance (see Figure \ref{fig:summary-sensitivity}).  

\paragraph{Dwarf spheroidal galaxies (dSphs).}
DSphs 
are among the best targets for DM searches for the following reasons (see, e.g., Ref. \cite{2015PhRvL.115w1301A}): {\tt (i)} the spread of their stellar velocities indicates very large mass to light ratios, hinting at a large DM content, up to O(1000) times more mass in DM than in visible matter; {\tt (ii)} their proximity compared to other targets of similar DM densities; {\tt (iii)} the fact that no signal is expected from these sources from standard astrophysical mechanisms. Thus, any detection would constitute a ``smoking gun'' for DM. Several tens of dSphs have been already discovered, with forecasts of a full census of tens more expected over the next years (for instance from searches with the Rubin Observatory \cite{2019arXiv190201055D}).
The CTA will point to the best available target in terms of expected DM signal: the one dSph with very high DM concentration and close proximity.
In the case of no detection, any DM limit would be considerably more robust than those obtained with other means of observation due to the lack of astrophysical foreground gamma-ray emission and the comparatively better determination of the underlying DM density profiles in this type of objects.
Figure~\ref{fig:summary-sensitivity}  shows the sensitivity of CTA to a WIMP annihilation signature as a function of WIMP mass for a CTA observation of the Sculptor dSph.

\paragraph{Galaxy clusters.}
Clusters of galaxies are the largest, most massive gravitationally bound systems in the Universe. 
Their high mass-to-light ratio and the presence of these systems at close distances, $z\lesssim 0.1$, from Earth, make galaxy clusters promising targets for decay and annihilation signals from DM. 
On one hand, the DM decay rate is directly proportional to the mass~\cite{Combet:2012tt}.
On the other hand, DM halos of galaxy clusters harbour an abundance of DM substructures, which are the result of the hierarchical structure formation in the $\Lambda$CDM paradigm.
These small substructures provide a substantial contribution to the expected overall DM annihilation signal, making galaxy clusters as competitive targets as the known dSphs \cite{2011JCAP...12..011S}.  
The improved angular resolution of CTA presents an extraordinary opportunity for studying these large, extended objects and to disentangle astrophysical  from DM signals.
CTA will observe different well-studied galaxy clusters 
like the Coma cluster \cite{Liang:2018mpc} and Fornax \cite{2012JCAP...07..017A}.
Yet, the biggest observation program will be carried out for the Perseus cluster \cite{Acciari:2018sjn}, with 300 hours of planned observation time.
Even in the worst-case scenario with no detection, it will be  possible to set stringent constraints in the WIMP phase space.
For example, previous bounds set by LAT observations for decaying DM could be improved by one to three orders of magnitude for DM mass above 1 TeV \cite{CTAConsortium:2018tzg}.

\chapter{Axions and axion-like particles}
An alternative candidate for DM is the axion and more generally axion-like particles (ALPs).
If these particles are sufficiently light and produced non-thermally in the Early Universe, they could constitute the entirety of cold DM (see, e.g.,  \cite{arias2012}).
Theory predicts that ALPs could oscillate into photons (and vice-versa) in the presence of external magnetic  fields \cite{raffelt1988}.

\begin{figure}[htb]
    \centering
    \includegraphics[width=0.8\linewidth]{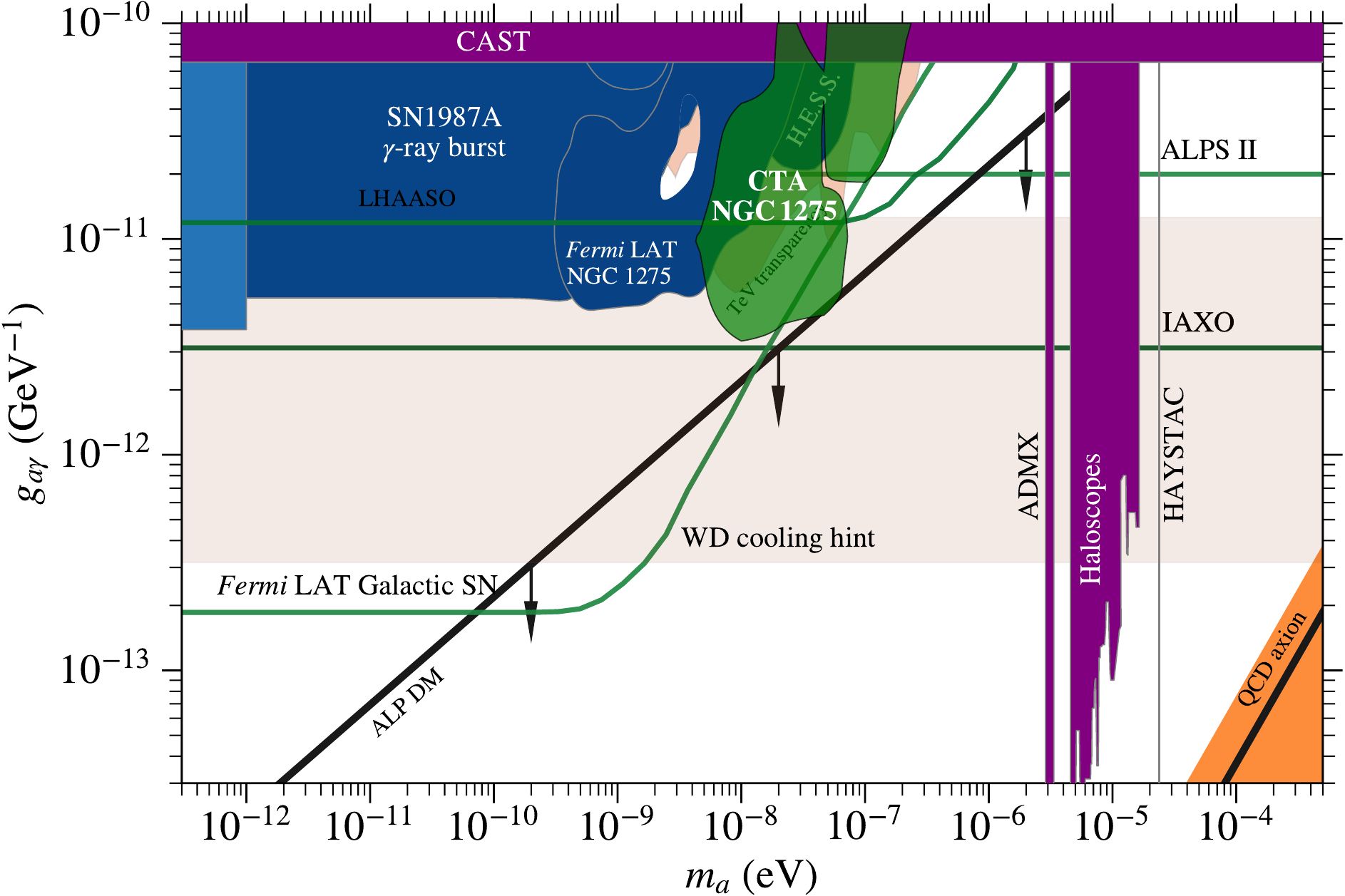}
    \caption{\label{fig:alps} The parameter space of ALPs given by their mass $m_a$ and coupling strength to photons $g_{a\gamma}$ taken from Ref. \cite{2021JCAP...02..048A}. 
    Current constraints are shown in blue and purple. Possible constraints from a single CTA observation of NGC~1275 are shown as a green filled region. These constraints start to reach the parameter space where ALPs could constitute the entire DM density (parameter space below the solid black line labeled ``ALP DM'').  
    }
\end{figure}

Photon-ALP conversions could take place in the various astrophysical magnetic fields and leave signatures detectable with CTA.
For example, due to the energy dependence of the photon-ALP oscillations, the usually smooth spectra of active galactic nuclei could be altered with distinctive oscillatory features, whose shape depends on the ALP mass, coupling strength to photons, and the morphology of the intervening magnetic fields. 
In a recent CTA Consortium publication it has been shown that a single CTA observation of NGC~1275 could test so-far unprobed regions in the ALP parameter space~\cite{2021JCAP...02..048A}, see Fig.~\ref{fig:alps}. 
Being the central galaxy of the Perseus cluster, which is known to harbor a strong magnetic field, NGC~1275 is among the prime candidates to search for such features. 

Furthermore, photon-ALP oscillations could reduce the apparent opacity of the Universe for very-high-energy gamma rays (see, e.g., Refs. \cite{deangelis2011,meyer2013, Kohri_2017}).
Instead of the production of electron-positron pairs in the interaction of gamma rays with background radiations fields, gamma rays could oscillate into ALPs with masses $\lesssim 10^{-6}\,$eV. 
If they reconvert back to photons close to Earth, the effective transparency would be increased in comparisons to the prediction of Standard Model physics. 
On the other hand, the decay of DM ALPs with masses around 1\,eV would contribute to the radiation fields that cause the opacity in the first place. 
This would lead to an increase of the background photon density and thereby an increased opacity for gamma rays \cite{2020A&A...633A..74K, 2020JCAP...03..064K}. 

With the planned observational program of monitoring active galaxies, following up on high emission states of such objects, and the extragalactic survey \cite{CTAConsortium:2018tzg}, CTA will have unprecedented sensitivity to search for such a non-standard gamma-ray opacity as well as the oscillatory features in gamma-ray spectra  \cite{meyer2014cta,2019MNRAS.487..123G,2012JCAP...07..017A}. 
Irrespective of the actual contribution of ALPs to the total DM density, CTA will be able to probe ALPs with masses below $\sim10^{-6}$\,eV, which is extremely difficult to achieve with astrophysical observations at other wavelengths. 

\chapter{Beyond Dark Matter: Searches for Lorentz Invariance Violation}
With the planned observational program of active galaxies and gamma-ray bursts, it will also be possible to probe the universality of Lorentz invariance with CTA.
Lorentz invariance violation (LIV) is predicted in theories beyond the Standard Model, such as models of quantum gravity or string theory (see, e.g., Refs. \cite{Amelino:2001,Alfaro:2005} and references therein).
The potential signatures of LIV in the gamma-ray channel include energy-dependent velocity of photons, photon decay, vacuum Cherenkov radiation, and pair-production threshold shifts \cite{1998Natur.393..763A,1999ApJ...518L..21K,Stecker:2001,Stecker:2003}. 
In a recent CTA Consortium publication the potential to test LIV-induced modifications of the pair-production threshold in gamma-ray interactions with  background radiation fields was assessed \cite{2021JCAP...02..048A}. 
A result of this modification is the suppression of pair-production above a certain energy and hence a reduction of the gamma-ray opacity. 
For a linear effect (in energy) of LIV on the photon dispersion, 
it was found that CTA observations of individual sources will probe energy scales at which LIV effects become important that are one order of magnitude above the Planck energy scale, $E_\mathrm{Pl}\sim10^{28}$\,eV.
This is comparable to current exclusions using the same signature but stacking many observations of currently operating Cherenkov telescopes \cite{2019PhRvD..99d3015L}. 
Thus, a stacking of CTA observations of multiple sources will provide unparalleled sensitivity for this effect. 

\bibliographystyle{ieeetr}
\bibliography{DMwcta}

\markboth{}{References}

\end{document}

%% file: author_list.tex

A.~Aguirre-Santaella (Instituto de F\'isica Te\'orica UAM-CSIC, Universidad Aut\'onoma de Madrid, Spain),
E.~Amato (INAF-Osservatorio Astrofisico di Arcetri, Italy),
E.O.~Anguner (Centre de Physique des Particules de Marseille, France),
L.A.~Antonelli (INAF-Osservatorio Astronomico di Roma, Italy),
Y.~Ascasibar (Universidad Aut\'onoma de Madrid, Spain),
C.~Bal\'azs (Monash University, Melbourne, Australia), 
G.~Beck (University of the Witwatersrand, South Africa), 
C.~Bigongiari (INAF-Osservatorio Astronomico di Roma, Italy),
J.~Bolmont (LPNHE, CNRS/IN2P3, Sorbonne University, France),
T.~Bringmann (University of Oslo),
A.M.~Brown (Durham University, U.K.),
M.G.~Burton (Armagh Observatory \& Planetarium, UK),
M.~Cardillo (INAF - IAPS, Roma, Italy)
S.~Chaty (University of Paris \& CEA, France),
G.~Cotter (University of Oxford, UK),
D.~della~Volpe (Universit\'e de Gen\`eve),
A.~Djannati-Ataï  (Université de Paris, CNRS/IN2P3, APC, France),
C.~Eckner (University of Nova Gorica),
G.~Emery, (Universit\'e de Gen\`eve),
E.~Fedorova (Taras Shevchenko National University of Kyiv),
M.~D.~Filipovic (Western Sydney University, Australia),
G.~Galanti (INAF - IASF Milano, Italy),
V.~Gammaldi (Instituto de F\'isica Te\'orica UAM-CSIC, Universidad Aut\'onoma de Madrid),
E.~M.~de Gouveia Dal Pino (IAG-USP - Universidade de São Paulo, Brazil),
J.~Granot (Open University of Israel, Israel),
J.G.~Green (INAF - Osservatorio Astronomico di Roma, Italy),
K.~Hayashi (National Institute of Technology, Ichinoseki College, Japan),
S.~Hern\'andez-Cadena (Instituto de F\'isica, Universidad Nacional Aut\'onoma de M\'exico),
N.~Hiroshima (RIKEN, Institute of Physical and Chemical Research, Japan), 
B.~Hnatyk (Taras Shevchenko National University of Kyiv),  
D.~Horan (LLR/Ecole Polytechnique, CNRS/IN2P3, France), 
M.~H\"utten (Max Planck Institute for Physics, Germany),
M. Jamrozy (Jagiellonian University, Poland),
A.~Lamastra (INAF - Osservatorio Astronomico di Roma, Italy),
J.-P.~Lenain (LPNHE, CNRS/IN2P3, Sorbonne Universit\'e, France),
E.~Lindfors (FINCA, University of Turku, Finland), 
I.~Liodakis (FINCA, University of Turku, Finland), 
S.~Lombardi (INAF-Osservatorio Astronomico di Roma, Italy),
F.~Longo (University and INFN, Trieste, Italy), 
F.~Lucarelli (INAF - Oss. Astr. di Roma and ASI-SSDC, Italy), 
K.~Kohri (Institute of Particle and Nuclear Studies, KEK, Japan), 
M.~Martinez (Institut de Física d’Altes Energies, IFAE-BIST, Barcelona, Spain),
H. Mart\'inez-Huerta (Departamento de F\'isica y Matemáticas, Universidad de Monterrey, M\'exico), 
D.~Mazin (ICRR, University of Tokyo, Japan and MPI for physics, Germany),
A.~Moralejo (Institut de Física d’Altes Energies, IFAE-BIST, Barcelona, Spain),
A.~Morselli (INFN Roma Tor Vergata, Italy),
C.G.~Mundell (University of Bath, UK), 
R.A.~Ong, (University of California, Los Angeles, USA), 
V.~Poireau (LAPP, CNRS/IN2P3, Univ. Savoie Mont Blanc, France), 
O.~Reimer (Innsbruck University, Institute for Astro- and Particle Physics, Austria),
J.~Rico (Institut de Fisica d'Altes Energies, Barcelona, Spain),
G.~Romeo (INAF - Osservatorio Astrofisico di Catania, Italy), 
P.~Romano (INAF, Osservatorio Astronomico di Brera), 
G.~Rowell (University of Adelaide, Australia), 
I.~Sadeh (DESY-Zeuthen, Germany),
M.A.~S\'anchez-Conde (Instituto de F\'isica Te\'orica UAM-CSIC, Universidad Aut\'onoma de Madrid, Spain),
F.G.~Saturni (INAF - Osservatorio Astronomico di Roma, Italy), 
O.~Sergijenko (Taras Shevchenko National University of Kyiv), 
H.~Sol (LUTH, Observatoire de Paris, CNRS, France), 
A.~Stamerra (INAF, Osservatorio Astronomico di Roma, Italy), 
Th.~Stolarczyk (AIM, CEA, CNRS, Université Paris-Saclay, Université de Paris, France),
F.~Tavecchio (INAF, Osservatorio Astronomico di Brera), 
S.~Vercellone (INAF, Osservatorio Astronomico di Brera), 
V.~Testa (INAF - Osservatorio Astronomico di Roma, Italy), 
L.~Tibaldo (IRAP - Universit\'e de Toulouse, CNRS, UPS, CNES, France), 
M.~Vecchi (Kapteyn Astronomical Institute, University of Groningen), 
A.~Viana (IFSC - Universidade de São Paulo,  Brazil), 
V.~Vitale (INFN - Roma Tor Vergata),
V.~Zhdanov (Taras Shevchenko National University of Kyiv)
on behalf of the CTA consortium